\documentstyle[preprint,eqsecnum,prd,aps,epsf,floats]{revtex}
\begin{document}

\newcommand{\be}{\begin{equation}}
\newcommand{\ee}{\end{equation}}
\newcommand{\nl}{\nonumber \\}
\newcommand{\q}{ {\cal Q} }
\newcommand{\qt}{ \tilde {\cal Q} }
\newcommand{\r}{ {\cal R} }
\newcommand{\rt}{ \tilde {\cal R} }
\newcommand{\oln}{\overline}

\newcommand{\drawsquare}[2]{\hbox{%
\rule{#2pt}{#1pt}\hskip-#2pt
\rule{#1pt}{#2pt}\hskip-#1pt
\rule[#1pt]{#1pt}{#2pt}}\rule[#1pt]{#2pt}{#2pt}\hskip-#2pt
\rule{#2pt}{#1pt}}

\newcommand{\Yfund}{\raisebox{-.5pt}{\drawsquare{6.5}{0.4}}}
\newcommand{\Ysymm}{\raisebox{-.5pt}{\drawsquare{6.5}{0.4}}\hskip-0.4pt%
        \raisebox{-.5pt}{\drawsquare{6.5}{0.4}}}
\newcommand{\Yasymm}{\raisebox{-3.5pt}{\drawsquare{6.5}{0.4}}\hskip-6.9pt%
        \raisebox{3pt}{\drawsquare{6.5}{0.4}}}
\newcommand{\Ythree}{\raisebox{-3.5pt}{\drawsquare{6.5}{0.4}}\hskip-6.9pt%
        \raisebox{3pt}{\drawsquare{6.5}{0.4}}\hskip-6.9pt
        \raisebox{9.5pt}{\drawsquare{6.5}{0.4}}}

\preprint{\vbox{\hbox{CALT-68-2125}
                \hbox{hep-th/9707071}  }}
 
\title{The Coulomb branch of $N=1$ supersymmetric $SU(N_c)\times SU(N_c)$
gauge theories }
 
\author{Martin Gremm}
 
\address{California Institute of Technology, Pasadena, CA 91125}

\maketitle 

\begin{abstract}
We analyze the low energy behavior of $N=1$ supersymmetric gauge theories
with $SU(N_c)\times SU(N_c)$ gauge group and a Landau-Ginzburg type
superpotential. These theories contain fundamentals transforming 
under one of the gauge groups as well as
bifundamental matter which transforms as a fundamental under each.
We obtain the parametrization of the gauge coupling on the Coulomb branch 
in terms of a hyperelliptic curve. The derivation of this curve involves
making use of Seiberg's duality for SQCD as well as the classical constraints
for $N_f=N_c+1$ and the quantum modified constraints for $N_f=N_c$.
\end{abstract}

\newpage

\section{Introduction}
Our understanding of the low energy behavior of supersymmetric gauge theories
has increased substantially over the last few years. Seiberg and Witten
found a complete solution for $N=2$, $N_c=2$ SQCD on the Coulomb branch \cite{sw}.
These results were later generalized to other gauge groups and to include
fundamental matter \cite{Curves}. The Coulomb branch is the segment of the
moduli space of vacua where the microscopic gauge group of rank $r$ is broken to
$U(1)^r$.
On the Coulomb branch the massless particles in the low energy theory
are the photons corresponding to the unbroken $U(1)$'s. These photons are described
by the Lagrangian
\begin{equation}
{\cal L}=\frac{1}{4\pi}{\rm Im} \int d^2\theta \tau_{ij} W^i W^j,
\end{equation}
where $\tau_{ij}$ is the matrix of the $U(1)$ gauge couplings. For $N=2$
theories this description
has to be supplemented by a kinetic energy term for the adjoint scalars
and at some points on the moduli space by terms describing particles that
go massless there. The effective Lagrangian can be expressed
as an integral of a holomorphic prepotential over half of $N=2$ superspace.
This provides a relation between the prepotential, the gauge couplings and
 the metric on the moduli space.
Thus, determining the gauge couplings as a function of the moduli
amounts to solving the low energy theory in that case.

In the $N=1$ case the low energy Lagrangian cannot be written in terms of 
a prepotential. Therefore, it is no longer sufficient to determine the 
gauge couplings as a function of the moduli in order to obtain a complete
solution of the low energy theory.
There is no simple relation between the $U(1)$ gauge couplings and the kinetic 
energy terms of the matter fields in the $N=1$ Lagrangian.
Nevertheless, it was shown in \cite{is} that in some cases the $U(1)$ gauge
couplings can be determined using the same methods as in the $N=2$ case.
A number of examples of such $N=1$ theories have been found
\cite{is,is2,anton,lg1,lg2,josh}. Only one of these examples \cite{is},
which was generalized in \cite{josh}, involves product gauge groups.
We provide a second such example here.

In both $N=1$ and $N=2$ theories, the matrix of $U(1)$ gauge couplings
$\tau_{ij}$ can be identified with the normalized period matrix of a 
Riemann surface. If the theories contain only fundamental matter,
this surface is usually hyperelliptic and
of genus $r$ where $r$ is the rank of the gauge group.
In these cases it can generally be determined uniquely using symmetry
and field theory arguments. In more complicated cases, solutions were 
obtained using D-brane configurations \cite{branes}.

In this paper we analyze $N=1$ supersymmetric gauge theories with
$SU(N_c)\times SU(N_c)$
gauge group and fundamental matter as well as bifundamental matter which
transforms as a fundamental under both gauge groups. These theories
have a Coulomb branch with an unbroken $U(1)^{N_c-1}$ gauge group
if a Landau-Ginzburg-type superpotential is added.
The values of the $U(1)$ gauge couplings can be parametrized in terms of
a hyperelliptic curve.

The theories we analyze here have some novel features.
In order to derive the curve it is necessary to consider
limits where one or the other $SU(N_c)$ is strongly coupled. Taking some
of these limits involves passing to a dual description of the strongly coupled
group. The duals involved are very similar to those found by Seiberg
\cite{dual}. For certain numbers of fundamental flavors the classical
constraints that arise in SQCD
for $N_f=N_c+1$ play a role. In the other examples of theories with product
gauge groups only the quantum modified constraints which arise for $N_f=N_c$
appeared. This is due to the fact that those theories did not contain matter
transforming as a fundamental under only one of the gauge groups.

The paper is organized as follows: In Section II we describe the
matter content and superpotential of the $SU(N_c)\times SU(N_c)$ theories.
We also derive the dual description and constraint equations that arise if one 
switches off one or the other gauge group. These results will be needed in
Section III to derive the curve for the $SU(N_c)\times SU(N_c)$ theories.
Concluding remarks can be found in Section IV.

\section{Preliminaries}

In this section we analyze various features of the theory with gauge group
$SU(N_c)_1\times SU(N_c)_2$ and the matter content given in
Table \ref{electric}.
\begin{center}
\begin{table}
\begin{tabular}{c|cc|ccccc}
 & $SU(N_c)_1$ & $SU(N_c)_2$ & $SU(N_f)_L$ & $SU(N_f)_R$ & $U(1)_B$ &
	$ U(1)_C$ & $U(1)_R$ \\ \hline
  $\q$ & $\Yfund$ & $1$ & $\Yfund$ & $1$ &  $1$ &  $0$ & $1$ \\
  $\qt$ & $\oln{\Yfund}$ & $1$ & $1$ & $\Yfund$ & $-1$ & $ 0$ & $1$ \\
  $\r$ & $\Yfund $ & $\oln{\Yfund} $ & $1$ & $1$ & $0$ & $1$ & $0$ \\
  $\rt$  & $\oln{\Yfund}$  & $\Yfund$ & $1$ & $1$ & $ 0$ & $-1$ &$ 0$ 
\end{tabular}
\caption{The matter content of the $SU(N_c)_1\times SU(N_c)_2$ gauge theory}
\label{electric}
\end{table}
\end{center}
The $N_f$ fields $\q$ and $\qt$ transform as
fundamentals under $SU(N_c)_1$ and are singlets under $SU(N_c)_2$.
The fields $\r$ and $\rt$ transform as fundamentals in one
and as antifundamentals in the other gauge group. 
The table also shows the nonanomalous assignment of the charges under the global
symmetry $SU(N_f)_L\times SU(N_f)_R\times U(1)_B \times U(1)_C\times U(1)_R$,
where $U(1)_R$ is an R-symmetry.  If we add a tree level superpotential of the
form
\be
\label{w}
W = \sum_{k=0}^l h^{(k)}_{ij} \qt^i ( \r\rt)^k \q^j,
\ee
where $i,j = 1,\ldots , N_f$, this theory has a Coulomb branch.
For $\q=\qt=0$ one can verify that the solution of the D-flatness conditions
for the $\r$ and $\rt$ fields has $N_c+1$ free parameters and that the vevs of
$\r$ and $\rt$ can be brought into diagonal form. Therefore, the low energy
theory has an unbroken $U(1)^{N_c-1}$ gauge group \cite{josh}.
There is also a Higgs branch with nonzero vevs for the quarks
on which the gauge group is broken completely. We will limit our discussion
to the Coulomb branch in this paper.

The Coulomb branch cannot be lifted by a dynamically generated superpotential,
since the nonanomalous $R$-charge assignment $R_{\cal R} = R_{\tilde{\cal R}}=0$
and $R_{\cal Q} =R_{\tilde{\cal Q}} =1$
requires any such superpotential to be quadratic in the quarks. The F-flatness
condition arising from any superpotential will automatically be satisfied on
the Coulomb branch \cite{anton} (but not necessarily on the Higgs branch).

The superpotential, Eq.~(\ref{w}), includes terms that make the theory
nonrenormalizable. It should be viewed as an effective field theory which is
defined below some scale $\Lambda$. We assume that all scales appearing in the 
effective theory are much smaller than $\Lambda$.
The dimensionful coefficients $h^{(k)}$ in the superpotential scale as
$1/\Lambda^{2k-1}$.

The low energy theory simplifies considerably if we take one of the two
gauge groups to be much more strongly coupled than the other, i.e.,
$\Lambda_1 \ll \Lambda_2$ or $\Lambda_2 \ll \Lambda_1$. These limits are 
analyzed most easily if we switch off the weakly coupled group, discuss the
resulting single gauge group theory without superpotential,
and then promote the $SU(N_c)$ that was switched off to a gauge symmetry again.
This is the procedure followed in, e.g.,
\cite{poppitz} to find dual descriptions for theories with product gauge groups.
Once a description of the theory in these limits is found, we can perturb it by adding the superpotential Eq.~(\ref{w}).
Other perturbations of the $SU(N_c)\times SU(N_c)$ theory we described above
were studied in \cite{leigh}.

If $SU(N_c)_1$
is switched off, the fields $\r$ and $\rt$ look like $N_c$
flavors of fundamentals from the point of view of $SU(N_c)_2$. The $SU(N_c)_2$
gauge theory with no superpotential is in the confining phase, i.e. the
low energy description should be in terms of the composite meson and baryon
fields $\Psi$ and $B,\tilde{B}$ made from $\r$ and $\rt$. These fields have to
satisfy the quantum modified constraint
\cite{dual}  
\be
\label{nfeqnc}
{\rm det} \Psi - B\tilde{B} = \Lambda_2^{2N_c}.
\ee
Here $\Lambda_2$ is the strong coupling scale of $SU(N_c)_2$.
Note that the meson $\Psi=\r\rt$  transforms as an adjoint plus a scalar under
$SU(N_c)_1$. In this limit, the superpotential, Eq.~(\ref{w}), takes the form
\be\label{wanton}
W = \sum_{k=0}^l h^{(k)}_{ij} \qt^i \Psi^k \q^j.
\ee
The Coulomb branch of this theory was discussed in \cite{anton}. At scales much
below $\Lambda_2$ no trace of the fact that $\Psi$ is a composite survives.
Therefore, one can follow the arguments of \cite{anton} to determine
that the superpotential is a relevant perturbation for $lN_f<2N_c$. We will
restrict our discussion in this paper to superpotentials that satisfy this
constraint.

Switching off $SU(N_c)_2$, we have a $SU(N_c)_1$ gauge theory with two types
of flavors in the fundamental representation. 
This theory with no superpotential is very similar to SQCD \cite{dual} with a
total of
$N_c+N_f$ flavors. We need to distinguish two cases: $N_f=1$ corresponding
to $N_f=N_c+1$ in \cite{dual} and $N_f>1$, in which case one expects the
theory to have a dual description \cite{dual}.

For $N_f=1$ the theory confines without chiral symmetry breaking, which
can be verified by computing the anomalies for the elementary particles listed
in Table \ref{electric} and for the composites in Table \ref{confined}.
\begin{table}\begin{center}
\begin{tabular}{c|cc|ccc}
 & $SU(N_c)_1$ & $SU(N_c)_2$ & $U(1)_B$ & $U(1)_C$ & $U(1)_R$ \\ \hline
  $M=\q\qt$ & $1$ & $1$ & $ 0$ &  $0$ & $2$ \\
  $P=\rt\q$     & $1$ & $\Yfund$       &  $1$ & $-1$ & $1$ \\
$\tilde{P}=\qt\r$ & $1$ & $\oln{\Yfund}$ & $-1$ &  $1$ & $1$ \\
  $\Psi=\rt\r$  & $1$ & $\rm{adj}+1$   &  $0$ &  $0$ & $0$ \\
\hline
  $B_0=\r^{N_c}$           & $1$ & $1$            &  $0$ &  $N_c$   & $0$ \\
$\tilde{B_0}=\rt^{N_c}$      & $1$ & $1$            &  $0$ & $-N_c$   & $0$ \\
  $B_1=\q\r^{N_c-1}$       & $1$ & $\Yfund$ &  $1$ & $N_c-1$  & $1$ \\
$\tilde{B_1}=\qt\rt^{N_c-1}$ & $1$ & $\oln{\Yfund}$       & $-1$ & $-N_c+1$ & $1$ \\
 \end{tabular}
\caption {The composites of the confining $SU(N_c)_1$ with $N_c+1$ flavors}
\label{confined}
\end{center}
\end{table}
As expected from the analysis in \cite{dual}, there are classical constraints
on the composite fields which cannot be modified quantum mechanically.
They follow as equations of motion from the superpotential
\be\label{classconstr}
W=\frac{1}{\Lambda_1^{2N_c-1}}\left(
	{\rm det} {\cal M} -  B_0  M \tilde{B}_0 - B_1 \Psi \tilde{B}_1 -
	B_0 P \tilde{B}_1 - B_1 \tilde{P} \tilde{B}_0 \right),
\ee
where
\be 
{\cal M} = \left( \begin{array}{c|c}
	M       & \tilde{P}    \\ \hline
	P & \Psi \\
  \end{array} \right).
\ee

For $N_f>1$ the theory has the necessary number of flavors to be in the duality
regime. However, the global symmetries of the theory under consideration here
differ from those of the theory discussed by Seiberg \cite{dual}. The dual
gauge group turns out to be $SU(N_f)$ as expected but the charge assignments
of the magnetic quarks differs from that in \cite{dual}. In Table
\ref{magnetic} we give the gauge group and particle content of the dual theory.
\begin{table}\begin{center}
 \begin{tabular}{c|cc|ccccc}
 & $SU(N_f)_1$ & $SU(N_c)_2$ & $SU(N_f)_L$ & $SU(N_f)_R$ & $U(1)_B$ & $U(1)_C$ & $U(1)_R$ \\ \hline
     $q$     & $\Yfund$       & $1$            & $\oln{\Yfund}$ & $1$ &
		 $0$ &  $\frac{N_c}{N_f}$ & $0$\\
  $\tilde q$ & $\oln{\Yfund}$ & $1$ & $1$ & $\oln{\Yfund}$ & $0$ &  $-\frac{N_c}{N_f}$ & $0$ \\
     $r$     & $\Yfund$       & $\Yfund$       & $1$            & $1$ &
		 $1$ & $ -1+\frac{N_c}{N_f}$ & $1$ \\
  $\tilde r$ & $\oln{\Yfund}$ & $\oln{\Yfund}$ & $1$            & $1$ &
		$-1$ & $1-\frac{N_c}{N_f}$ & $1$ \\
$M=\q\qt$    &  $1$           &  $1$           & $\Yfund$       & $\Yfund$ &
		 $0$ & $0$                 & $2$ \\
$P=\rt\q$    &  $1$           &  $\Yfund$      & $\Yfund$       & $1$ &
		 $1$ & $-1$                & $1$ \\
$\tilde P=\r\qt$  &  $1$      & $\oln{\Yfund}$ & $1$            & $\Yfund$ &
		$-1$ &  $1$                & $1$ \\
$\Psi=\rt\r$ &  $1$           & $\rm{adj}+1$ & $1$ & $1$ &
		$0$ & $0$ & $0$ \\
\end{tabular}
\caption{The matter content of the dual theory}
\label{magnetic}
\end{center}
\end{table}
One can verify that the anomalies of the global symmetries match. On the
dual side we have to add a superpotential of the form
\be\label{wmag}
W= M q \tilde q + P q \tilde r + \tilde P \tilde q r + \Psi r \tilde r,
\ee
to remove the bilinears of dual quarks from the chiral ring.
The matching of baryons works as follows
\be
\q^p \r^{N_c-p} \rightarrow r^p q^{N_f-p}.
\ee
Note that the dual theory has an $SU(N_c)$ global symmetry which can be
gauged.

The constraints and the dual given here will be needed in the next section to 
discuss the limiting behavior of the curve which describes the Coulomb branch
of the $SU(N_c)\times SU(N_c)$ theory.

\section{The curve }

The theory with gauge group $SU(N_c)_1\times SU(N_c)_2$,
the superpotential given in Eq.~(\ref{w}), and the matter content in Table
\ref{electric} has two limits in which it reduces to theories for which the
curve describing the gauge couplings on the coulomb branch is known.
We can use these limits to constrain the curve for the theory we are considering
here. If one integrates out all fundamentals, the resulting
curve has to reproduce the curve for the $SU(N_c)\times SU(N_c)$ case given in
\cite{josh}. In the limit $\Lambda_2 \gg \Lambda_1$ the fields 
transforming under the $SU(N_c)_1$ gauge group are $N_f$ fundamentals and an
adjoint with the superpotential Eq.~(\ref{wanton}). This is the 
theory discussed in \cite{anton}, so in this limit the curve has to agree with
the one given there. The analysis of the limit $\Lambda_1 \gg \Lambda_2$ 
is somewhat more involved but it will turn out that this limit also yields
a theory of the type studied in \cite{anton}.

From the solution of the D-flatness conditions \cite{josh} we know that $N_c-1$
$U(1)$'s remain unbroken on the Coulomb branch. It is convenient to define 
\be\label{psi}
\Phi = \Psi - \frac{1}{N_c} {\rm Tr} \Psi =
	\r\rt - \frac{1}{N_c} {\rm Tr} \r\rt,
\ee
which transforms as an adjoint under $SU(N_c)_1$.
The diagonal form $\Phi = {\rm diag}(\phi_1,\ldots,\phi_{N_c})$ can be used
to define the gauge invariant symmetric polynomial
\be\label{defs} 
\prod_{j=1}^{N_c} (x-\phi_j)=\sum_{j=0}^{N_c} s_j x^{N_c-j} .
\ee
In terms of these variables the curve reads
\be\label{curve}
y^2 = \left[ \sum_{j=0}^{N_c} s_j x^{N_c-j} + (-1)^{N_c}\left(
	\Lambda_2^{2N_c}+ \Lambda_1^{2N_c-N_f} \det h^{(0)} \right)
	\right]^2-4\Lambda_2^{2N_c}\Lambda_1^{2N_c-N_f}
		\det\sum_{j=0}^lh^{(j)}x^j.
\ee
There could be other terms in this curve which are allowed by the symmetries
of the theory but
they can be excluded on the basis of the limits we discuss below.

If one takes all $h^{(i)},i\ne 0$ to vanish and the entries in the 
mass matrix $h^{(0)}$
to be large, one can integrate out all flavors of quarks. In this case, the
curve has to reproduce that given in \cite{josh}. It is a simple matter
to check that this is in fact the case.

The solution of the D-flatness conditions implies that the vev of the 
fields $\r$ and $\rt$ can be brought into diagonal form \cite{josh}.
Giving $\r$ a large diagonal vev, i.e., $\r={\rm diag}(v,\ldots,v)$,
breaks the product gauge group to its diagonal subgroup $SU(N_c)_D$. 
Both bifundamentals decompose into an adjoint and a singlet under $SU(N_c)_D$,
and the quarks $\q$ and $\qt$ transform as fundamentals.
One of the adjoints is eaten by the Higgs mechanism.
Rewriting the superpotential in terms of the uneaten adjoint
$\Psi_{\tilde{\cal R}} = \rt - 1/N_c {\rm Tr} \rt$, we find
\be
W = \sum_{k=0}^l h^{(k)}_D \qt \Psi_{\tilde{\cal R}}^k \q,
\ee
where $h^{(k)}_D=h^{(k)} v^{k}$. This theory has the same matter content and
superpotential as the theory of \cite{anton}. Thus we have
to recover the curve given there in this limit. The matching
relation for the strong coupling scales determines
\be 
\Lambda_1^{2N_c-N_f} \Lambda_2^{2N_c} = v^{2N_c}\Lambda_D^{2N_c-N_f}.
\ee
Finally, we need to rewrite the gauge invariant polynomials in terms of the
components of $\Psi_{\tilde{\cal R}}$
\be 
s_j = v^j s_j^{(D)}.
\ee
Substituting these expressions into the curve Eq.~(\ref{curve}), rescaling
$x\to x/v, \, y \to y/v^{N_c}$ and neglecting subleading terms gives
\be
y^2 = \left[ \sum_{j=0}^{N_c} s^{(D)}_j x^{N_c-j} \right]^2
	-4\Lambda_D^{2N_c-N_f} \det\sum_{j=0}^lh_D^{(j)}x^j,
\ee
which agrees with \cite{anton}.

In the limit $\Lambda_2 \gg \Lambda_1$ the second gauge group confines
and we need to rewrite the curve in terms of the composite degrees of freedom.
The composites fields are related by the constraint Eq.~(\ref{nfeqnc}). This 
constraint can be incorporated by shifting the gauge invariant polynomial
$s_{N_c}\rightarrow s_{N_c} - (-1)^{N_c} \Lambda_2^{2N_c}$ \cite{josh}.
We also need to rescale the $s_j$ to give them the canonical mass dimensions
\begin{eqnarray}\label{rfields} 
s_j= \mu^j s_j^{(1)},\,\, j=1,\ldots,N_c-1 \nl
s_{N_c} = \mu^{N_c} s_{N_c}^{(1)} - (-1)^{N_c} \Lambda_2^{2N_c},
\end{eqnarray}
where $\mu$ is some as yet undetermined mass scale.
The polynomials on the right hand side have the correct mass dimensions and
incorporate the constraint automatically.
Rescaling the field $\Psi$ in the 
superpotential, Eq.~(\ref{w}), to give it the right mass dimension requires 
a simultaneous redefinition of the coefficients
$h^{(k)} \mu^k = h^{(k)}_L$. The matching relation for the 
strong coupling scales reads
\be\label{match}
\Lambda_1^{2N_c-N_f} \Lambda_2^{2N_c} = \mu^{2N_c} \Lambda_L^{2N_c-N_f}  .
\ee
Substituting this, the rescaled coefficients, and the rescaled fields,
Eq.~(\ref{rfields}), into the curve Eq.~(\ref{curve}), one finds
\be\label{limit1}
y^2 = \left[ \sum_{j=0}^{N_c} s_j^{(1)} x^{N_c-j} \right]^2
	-4\Lambda_L^{2N_c-N_f} \det\sum_{j=0}^lh^{(j)}_L x^j,
\ee
which agrees with the curve given in \cite{anton}.
In this expression we rescaled $x\to x/\mu$, $y\to y/\mu^{N_c}$, set
$\mu = \Lambda_2$ and neglected an irrelevant subleading piece of the form
$\Lambda_1^{2N_c-N_f} \det h^{(0)} /\Lambda_2^{2N_c}$. 
Note that the quantum piece of this curve vanishes whenever one of the quarks
becomes massless classically. This can be seen as follows:
The superpotential has the structure of a mass term for the quarks.
Choosing a basis in which $\Psi = {\rm diag}(\phi_1,\ldots,\phi_{N_c})$,
the superpotential takes the form 
\be 
W=\sum_{k=0}^l h^{(k)}_{ij} \sum_\alpha \phi_\alpha^k \q^{i\alpha}
	\qt^j_{\alpha}.
\ee
Whenever 
\be
\det \sum_{k=0}^l h^{(k)}_{ij} \phi_\alpha^k  = 0
\ee
is satisfied, at least one of the quarks charged under the corresponding $U(1)$
becomes massless. This condition constrains the quantum piece of the curve
\cite{anton}. After rescaling the composite field and the coefficients $h^{(k)}$
in the superpotential, the quantum piece is given by
\be 
\Lambda^{2N_c-N_f}_L\det \sum_{k=0}^l h^{(k)}_L x^k ,
\ee
in agreement with Eq.~(\ref{limit1}).
We will use similar considerations to check the curve Eq.~(\ref{curve}) 
in other limits.

The limit $\Lambda_1 \gg \Lambda_2$ is more complicated because, from the
point of view of $SU(N_c)_1$, there are more flavors than colors. In order to 
determine whether the curve describes this limit correctly, we have to analyze 
the description of the low energy physics in terms  of the composite degrees
of freedom. The composites $\Psi$ and $\Phi$ have to be
redefined by switching the order
of $\r$ and $\rt$ in Eq.~(\ref{psi}), so that they transform as adjoints
under $SU(N_c)_2$. However, this does not change the values of the gauge
invariant polynomials $s_j$. Thus the classical piece of the curve is unchanged.
We can use the techniques of \cite{anton} to find the curve
corresponding to the description in terms of the composites
and compare it to the appropriate limit of the curve Eq.~(\ref{curve}).

If there is only one flavor of the quarks $\q$ and $\qt$, the first gauge
group sees a total of $N_c+1$ flavors. It is in the confining phase and 
the composite degrees of freedom have to satisfy the constraints following from
Eq.~(\ref{classconstr}).  In order to discuss the theory in this limit,
we need to reexpress the tree level superpotential, Eq.~(\ref{w}),
in terms of the confined composites and add the superpotential,
Eq.~(\ref{classconstr}), to incorporate
the constraint on these fields. Using the operator maps in Table \ref{confined},
we find
\be\label{wconv}
W = h^{(0)} M+\sum_{k=1}^l h^{(k)} \tilde{P} \Psi^{k-1} P
	-\frac{1}{\Lambda_1^{2N_c-1}} \left(
{\rm det} {\cal M} -   B_0  M \tilde{B}_0 - B_1 \Psi \tilde{B}_1 -
	B_0 P \tilde{B}_1 - B_1 \tilde{P} \tilde{B}_0\right).
\ee
The matter content of the $SU(N_c)_2$ gauge theory consists of the singlets
$M$ and $B_0,\,\tilde{B}_0$, two flavors of quarks $P,\,\tilde{P}$
and $B_1,\,\tilde{B}_1$ and the adjoint $\Psi$. The
singlets do not take part in the gauge dynamics but $B_0$ and $\tilde{B}_0$
serve as off-diagonal mass terms for the two flavors of fundamentals.
Except for the presence of the singlet $M$, this limit of the theory is similar
to the theory considered in \cite{anton}.
We can repeat the derivation given there to find the curve for this theory.
To do so, we need to determine the classical condition for the quarks to become
massless. 
The determinant in the superpotential, Eq.~(\ref{wconv}), can be expanded using
the diagonal representation of $\Psi$,
\be\label{detm}
{\rm det} {\cal M} = \left( M - \sum_{\alpha=1}^{N_c}
\frac{P_\alpha \tilde{P}_\alpha}{\phi_\alpha} \right) \prod_{\beta=1}^{N_c} \phi_\beta.
\ee
Substituting this into the superpotential, the equation of motion for
$M$ requires
\be\label{constr} 
\det \Psi - B_0 \tilde{B}_0 = h^{(0)} \Lambda_1^{2N_c-1},
\ee
where we have reexpressed the product of the $\phi_i$ as a determinant.
Note that this constraint involves only the composites made from the fields
$\r$ and $\rt$.
All the terms in the superpotential, Eq.~(\ref{wconv}), that involve fields
which transform as fundamentals under $SU(N_c)_2$ have the structure of
mass terms. We could analyze those, using the constraint Eq.~(\ref{constr}),
to determine where the composite quarks $P$ and $B_1$ become massless. 
However, it is much easier to impose the constraint by integrating out $M$ and
analyze the resulting superpotential. This yields
\be
W = \sum_{k=0}^l h^{(k)} \sum_{\alpha}
	P_\alpha \tilde{P}^\alpha \phi_\alpha^{k-1}
	+\frac{1}{\Lambda_1^{2N_c-1}}\left(
	B_1 \Psi \tilde{B}_1 + B_0 P \tilde{B}_1 + B_1 \tilde{P} \tilde{B}_0+
	\frac{P_\alpha \tilde{P}_\alpha}{\phi_\alpha} B_0 \tilde{B}_0
	\right),
\ee
which has the form of a mass term for the two flavors of quarks. By writing
these mass terms as a matrix in flavor space and requiring that its 
determinant vanishes, we find that at least one quark will become massless
if 
\be\label{massless}
\sum_{k=0}^l h^{(k)} \phi_\alpha^k \ = 0.
\ee
This implies that the quantum piece of the curve in this limit should be
proportional to
\be\label{qmpiece}
\Lambda_L^{2N_c-2}\sum_{j=0}^lh^{(j)}_L x^j
\ee
after rescaling the composite fields and the coefficients $h^{(k)}$ in the
superpotential.

In order to find the curve in the limit $\Lambda_1 \gg \Lambda_2$, we need
to rescale the gauge invariant polynomials and shift the highest one 
according to the constraint Eq.~(\ref{constr})
\begin{eqnarray}\label{rfields2} 
s_j= \mu^j s_j^{(2)}, j=1,\ldots,N_c-1 \nl
s_{N_c} = \mu^{N_c} s_{N_c}^{(2)} - (-1)^{N_c} h^{(0)} \Lambda_2^{2N_c-1}.
\end{eqnarray}
Rescaling the composites in the superpotential, Eq.~(\ref{wconv}), to give them
the canonical mass dimension one requires that we define
$h^{(k)}_L = h^{(k)} \mu^{k+1}$. Finally, the matching condition
for the strong coupling scales reads
\be\label{match2}
\Lambda_1^{2N_c-1} \Lambda_2^{2N_c} = \mu^{2N_c+1}\Lambda_L^{2N_c-2}  .
\ee
Substituting this and the rescaled polynomials, Eqs.~(\ref{rfields2}),
into the curve, Eq.~(\ref{curve}), gives
\be
y^2 = \left[ \sum_{j=0}^{N_c} s_j^{(2)} x^{N_c-j} \right]^2
	-4\Lambda_L^{2N_c-2}\sum_{j=0}^lh^{(j)}_L x^j,
\ee
where we rescaled $x\to x/\mu$, $y\to y/\mu^{N_c}$, set $\mu = \Lambda_1$
and neglected subleading terms. The quantum piece of this curve agrees 
with Eq.~(\ref{qmpiece}), i.e., the curve describes this limit correctly.

Finally, we have to check that the curve, Eq.~(\ref{curve}), gives the correct
description in the limit $\Lambda_1 \gg \Lambda_2$ for $N_f>1$. In this 
case, the $SU(N_c)_1$ gauge theory is in the dual phase if $SU(N_c)_2$
is switched off. In order to describe the low energy physics 
in this limit, we have to pass to the dual description.
The operators in the tree level superpotential are mapped to operators on
the magnetic side
according to Table \ref{magnetic}. We also need to add the superpotential,
Eq.~(\ref{wmag}), to eliminate gauge invariant combinations of the dual
quarks from the chiral ring. This results in the superpotential
\be\label{wmagtot} 
W = \mu h^{(0)}M + \sum_{k=1}^l \mu^{k+1} h^{(k)} P \Psi^{k-1} \tilde{P}
	+M q \tilde q + P q \tilde r + \tilde P \tilde q r + \Psi r \tilde r,
\ee
where we have inserted a scale $\mu$ in some of the terms to correct the mass
dimensions. This is necessary because we take the mesons $M$ and $P,\,\tilde{P}$
to have mass dimension one.
The mass term for the meson $M$ forces the quarks $q$ and $\tilde{q}$ to
acquire a vev, which higgses the dual $SU(N_f)$ gauge group completely for 
generic values of $h^{(0)}$. There are $2N_f$ flavors of quarks which transform
as fundamentals under $SU(N_c)_2$:
$N_f$ magnetic bifundamentals and $N_f$ mesons
$P$ and $\tilde{P}$. All terms in the superpotential, Eq.~(\ref{wmagtot}), except
those involving $M$ have the form of mass terms for the $2N_f$ flavors of
quarks. 
Again we must determine where these go massless classically, because this
determines the quantum piece of the curve.
Using $\Psi = {\rm diag}(\phi_1,\ldots,\phi_{N_c})$, we can rewrite the
mass terms as a matrix in flavor space and require that its determinant
vanishes. This yields the condition
\be
\det \sum_{k=0}^l \mu^{k+1} h^{(k)}_{ij} \phi_\alpha^k  = 0
\ee
on the vev of the adjoint after substituting $q\tilde{q}=-\mu h^{(0)}$.
We can now repeat the analysis of \cite{anton} to determine that the quantum
piece of the curve is proportional to
\be 
\Lambda_L^{2N_c-2N_f}\det \sum_{k=0}^l h^{(k)}_L x^k,
\ee
where we defined $h^{(k)}_L = \mu^{k+1} h^{(k)}$.
The quantum modified constraint
on the mesons and baryons made from the fields $\r$ and $\rt$,
\be 
\det \Psi - B_0 \tilde{B}_0 = \det h^{(0)} \Lambda_1^{2N_c-N_f},
\ee
can be obtained from the matching relations for the strong coupling scales
as one integrates in a flavor of $\q$. We can now repeat the same analysis
as in the confining case with the obvious modifications of Eqs.~(\ref{rfields2})
and the matching conditions for the strong coupling scales Eq.~(\ref{match2}).
Taking $\mu = \Lambda_1$, we find 
\be
y^2 = \left[ \sum_{j=0}^{N_c} s_j^{(2)} x^{N_c-j} \right]^2
	-4\Lambda_L^{2N_c-2N_f}\det\sum_{j=0}^lh^{(j)}_Lx^j,
\ee
which agrees with the curve obtained along the lines of \cite{anton} for the
case we consider here.
This concludes the checks on the curve given in Eq.~(\ref{curve}). 

\section{conclusion}

We have investigated the Coulomb branch of $SU(N_c)\times SU(N_c)$
gauge theories with fundamental and bifundamental matter and a 
Landau-Ginzburg type superpotential. In order to discuss the behavior of these
theories in the limit that one or the other gauge group is strongly
coupled, it is necessary to use Seiberg's results on confinement in $SU(N_c)$
theories with $N_f=N_c$ and $N_f=N_c+1$ as well as a dual description
for $N_f>N_c+1$. We found the curve that parametrizes the the gauge couplings 
of the unbroken $U(1)$'s and 
demonstrated that it reproduces known results in four limits. The product gauge
group can be broken to its diagonal subgroup, in which case we have to recover 
the curve given in \cite{anton}.
If all flavors are integrated out, we obtain the curve of
\cite{josh}. For both $N_f=1$ and $N_f>1$, the theory presented here 
reduces to theories considered in \cite{anton} if the limit $\Lambda_1 \gg
\Lambda_2$ or $\Lambda_2 \gg \Lambda_1$ is taken. In all of these cases,
we recover the curves given in \cite{anton}.
While this method of finding the curve is certainly
not rigorous, the evidence we have presented here strongly suggests that
our curve is the correct description of the $U(1)$ gauge couplings on the
Coulomb branch.

The curve can be used to analyze which particles go massless at its
singularities. Doing this explicitly for large values of $N_c$ is very
cumbersome. For $N_c=2$ such an analysis reproduces the results in
\cite{anton} except that one has to identify the variable $u$ in that paper
with $s_2+\Lambda_2^4+\Lambda_1^3h^{(0)}$. This shifts the location of the
singularities by a finite amount. The curve has singularities corresponding
to a monopole or a dyon going massless as well as singularities where the
quarks go massless. One can find a number of inequivalent
superconformal fixed points by tuning the coefficients $h^{(k)}$
such that some of the singularities corresponding to mutually nonlocal particles
collide. Such fixed points exist for $N_c \ge 2$, $N_f \ge 1$ and $l \ge 1$.
They are the $N=1$ analog (in the sense of \cite{lg2}) 
of the $N=2$ fixed points analyzed in \cite{adpoints}.

\acknowledgments

It is a pleasure to thank Lisa Randall, Joshua Erlich and Anton Kapustin
for several very helpful conversations and Eric Westphal for many useful
comments on the manuscript.
This work was supported in part by the U.S.\ Dept.\ of Energy under Grant no.\
DE-FG03-92-ER~40701.

{\tighten

}


\begin{references}

\bibitem{sw}
N. Seiberg and E. Witten, Nucl. Phys. B426, 19 (1994); Nucl. Phys. B431, 484 (1994).

\bibitem{Curves}
P.C. Argyres and A.E. Faraggi, Phys. Rev. Lett. 74, 3931 (1995);
A. Klemm, W. Lerche, S. Theisen, and S. Yankielowicz, Phys. Lett B344, 169 (1995);
U.H. Danielsson and B. Sundborg, Phys. Lett B358, 273 (1995);
A. Brandhuber and K. Landsteiner, Phys. Lett B358, 73 (1995);
P.C. Argyres, M.R. Plesser, and A.D. Shapere, Phys. Rev. Lett 75, 1699 (1995);
A. Hanany and Y. Oz, Nucl. Phys. B452, 283 (1995);
P.C. Argyres and A.D. Shapere, Nucl. Phys. B461, 437 (1996);
K. Landsteiner, J.M. Pierre, S.B. Giddings, Phys. Rev. D55, 2367 (1997).

\bibitem{is}
K. Intriligator and N.Seiberg, Nucl. Phys. B431, 551 (1995).

\bibitem{is2}
K. Intriligator and N.Seiberg, Nucl. Phys. B444, 125 (1995).

\bibitem{anton}
A. Kapustin, Phys. Lett. B398, 104 (1997).

\bibitem{lg1}
T. Kitao, S. Terashima, and S. Yang, Phys. Lett. B399, 75 (1997).

\bibitem{lg2}
A. Giveon, A. Pelc, and E. Rabinovici, hep-th/9701045.

\bibitem{josh}
C. Csaki, J. Erlich, D. Freedman, and W. Skiba, hep-th/9704067.

\bibitem{branes}
E. Witten, hep-th/9703166;
K. Landsteiner, E. Lopez, and D. Lowe, hep-th/9705199.

\bibitem{dual}
N. Seiberg, Phys. Rev. D49, 6857 (1994); Nucl. Phys. B435, 129 (1995).

\bibitem{poppitz}
E. Poppitz, Y. Shadmi, and S. Trivedi, Nucl. Phys. B480, 125 (1996).

\bibitem{leigh}
K. Intriligator, R.G. Leigh, and M.J. Strassler, Nucl. Phys. B456, 567 (1995).

\bibitem{adpoints}
P.C. Argyres and M.R. Douglas, Nucl. Phys. B448, 93 (1995);
P.C. Argyres, M.R. Plesser, N. Seiberg, and E. Witten, Nucl. Phys. B461, 71 (1996).

\end{references}
\end{document}